\documentclass[a4paper,10pt]{article}
\pdfoutput=1

\usepackage{devanagari}
\newcommand{\Tr}{\operatorname{Tr}}

\newcommand{\Psec}{P_{\rm sec}}
\newcommand{\Rcoll}{R_{\rm coll}}

\def\as(#1){{\alpha_{\rm s}^{\,#1}}}

\usepackage[T1]{fontenc} % if needed

\usepackage{
graphicx,
hyperref,
amsmath,
amssymb,
charter,
xcolor,
ifluatex,
longtable,
booktabs,
multirow,
bbold,cancel}

\usepackage{comment}

\usepackage{tikz}
\usepackage{tikz-feynman}
\tikzfeynmanset{compat=1.1.0}

\usepackage{multirow}

\usepackage{float}
\usepackage{colortbl}

\definecolor{Gray}{gray}{0.95}
\definecolor{RGray}{gray}{0.85}
\definecolor{CGray}{gray}{0.92}

\usepackage{multicol}
\definecolor{tit}{rgb}{0.1,0.2,0.4}
\definecolor{blus}{cmyk}{1,1,0,0.6}
\definecolor{verde}{cmyk}{0.92,0,0.59,0.25}

\usetikzlibrary{matrix}

\usepackage{tipa}
\usepackage{amsmath,amssymb,amsfonts, bm}
\usepackage{dsfont}
\usepackage{epsfig}
\usepackage{graphicx}
\usepackage{slashed}
\usepackage{color}

\usepackage{caption}
\usepackage{subcaption}
\usepackage{slashed} 

\usepackage{adjustbox}
\usepackage{commath}
\usepackage{calc}

\usepackage{sidecap}

\newcommand{\beq}{\begin{equation}}
\newcommand{\eeq}{\end{equation}}

\usepackage{array}

\makeatletter
\newcommand*{\rom}[1]{\expandafter\@slowromancap\romannumeral #1@}
\makeatother

\usepackage{float}
\usepackage{caption}

\begin{document}

\allowdisplaybreaks
\vspace*{-2.5cm}
\begin{flushright}
{\small
IIT-BHU
}
\end{flushright}

\vspace{2cm}

\begin{center}
{\LARGE \bf \color{tit} Naturally Light Composite Higgs as a Protected Collective Eigenmode   }\\[1cm]

{\large\bf Gauhar Abbas$^{a}$\footnote{email: gauhar.phy@itbhu.ac.in} }  
\\[7mm]
{\it $^a$ } {\em Department of Physics, Indian Institute of Technology (BHU), Varanasi 221005, India}\\[3mm]
%{\it $^b$  } {\em }\\[3mm]

\vspace{1cm}
{\large\bf\color{blus} Abstract}
\begin{quote}
Standard routes to a light composite Higgs either rely on tuning a
single channel near criticality or protect a pseudo-Nambu--Goldstone
coordinate of a coset. We
introduce a third mechanism class in which the protected object is an
\emph{eigenvalue} of the renormalized multi-operator scalar kernel of the
strong sector. Two basis-invariant diagnostics, a sector participation number
$\Psec$ and a sector gap ratio $\Rcoll$, identify collective lightness, but
they cannot distinguish an accidental small determinant from a protected zero
mode; the missing discriminator is the microscopic sensitivity
$\Delta_g=\left|\partial\ln|m_H^2|/\partial\ln g\right|$. A protected
collective Higgs is defined by $\Rcoll\gg1$, $\Psec>1$, and
$\Delta_g=\mathcal{O}(1)$. We prove that this class is nonempty. A rank-one
TC--DTC locking invariant forbids tree-level aligned curvature, while
universal vectorlike DQCD bridge fermions, massless in the microscopic
Lagrangian but acquiring a common DQCD constituent mass, obey
$\partial_h^2\sum_A\Tr\mathcal{M}_A^2\big|_0=0$. The aligned scalar is
therefore lifted only at joint two-spurion order,
$m^2_{\mathrm{br}}=-(d_X/2\pi^2)g_T^2g_D^2(f_B^4/f_H^2)L_X$, giving
$\Delta_{g_T}=\Delta_{g_D}=2$ at leading logarithmic order. The same DTC
topology admits a collective top completion and a vector-decoupling route to
reducing the positive technicolor contribution to $S$. The mechanism is
falsifiable by sector-restricted lattice spectroscopy and coupling-response
scans.

\end{quote}

\thispagestyle{empty}
\end{center}

\begin{quote}
{\large\noindent\color{blus} 
}

\end{quote}

\newpage
\setcounter{footnote}{0}
%\maketitle
%\flushbottom

\def\ca{{C^{}_{\!A}}}
\def\cf{{C^{}_F}}

\textit{Motivation.}---Known routes to a light composite Higgs fall into two
familiar classes. In Nambu--Jona-Lasinio (NJL)-like dynamics, a scalar becomes
light as a single channel approaches criticality~\cite{Nambu:1961tp,
Miransky:1988xi,Bardeen:1989ds}; in pseudo-Nambu--Goldstone (pNGB)
constructions, lightness follows from a coset shift
symmetry~\cite{Kaplan:1983fs,Contino:2010rs}. The first is a critical tuning,
the second a field-coordinate protection. A generic strongly coupled theory,
technicolor-like~\cite{Weinberg:1975gm,Susskind:1978ms,Cacciapaglia:2020kgq}
or otherwise, contains many scalar operators with identical unbroken quantum
numbers, so the physical Higgs is an eigenmode of a renormalized scalar
kernel, not a chosen channel. We propose a third mechanism: a \emph{protected
collective eigenmode}, light only in the full coupled kernel and stable
against microscopic coupling variations.

The result has three parts. First, spectral collectivity is a diagnostic but
not a naturalness mechanism; the basis-invariant pair $(\Rcoll,\Psec)$ must be
supplemented by the response exponent
\begin{equation}
\Rcoll\gg1,\qquad \Psec>1,\qquad \Delta_g=\mathcal{O}(1).
\label{eq:trichotomy}
\end{equation}
Second, a rank-one TC--DTC locking invariant plus a universal DQCD bridge
realizes Eq.~(\ref{eq:trichotomy}). Third, the same DTC topology admits a
collective top completion and lattice falsification.

\textit{Metric-covariant scalar kernel.}---Let $\mathcal{O}_r$ be renormalized
scalar operators with the same unbroken quantum numbers, and let $\phi_r$
denote the corresponding composite coordinates. Around a translationally
invariant vacuum the Euclidean quadratic Legendre kernel is
\begin{equation}
\begin{aligned}
\Gamma^{(2)}_E(Q^2) &= H + Q^2 Z + Q^4 Y(Q^2),\\
H&=H^\dagger,\qquad Z=Z^\dagger>0,
\end{aligned}
\label{eq:Gamma2}
\end{equation}
with $Y$ bounded near the origin. Canonical normalization gives
\begin{equation}
C = Z^{-1/2} H Z^{-1/2},\qquad H w_a = \mu_a^2 Z w_a,
\label{eq:gevp}
\end{equation}
with $w_a^\dagger Z w_b=\delta_{ab}$. The $Z$ metric is essential: operator
rescalings or rotations otherwise change any statement about which sector the
Higgs ``mostly'' occupies.

\begin{figure*}[t]
\includegraphics[width=\textwidth]{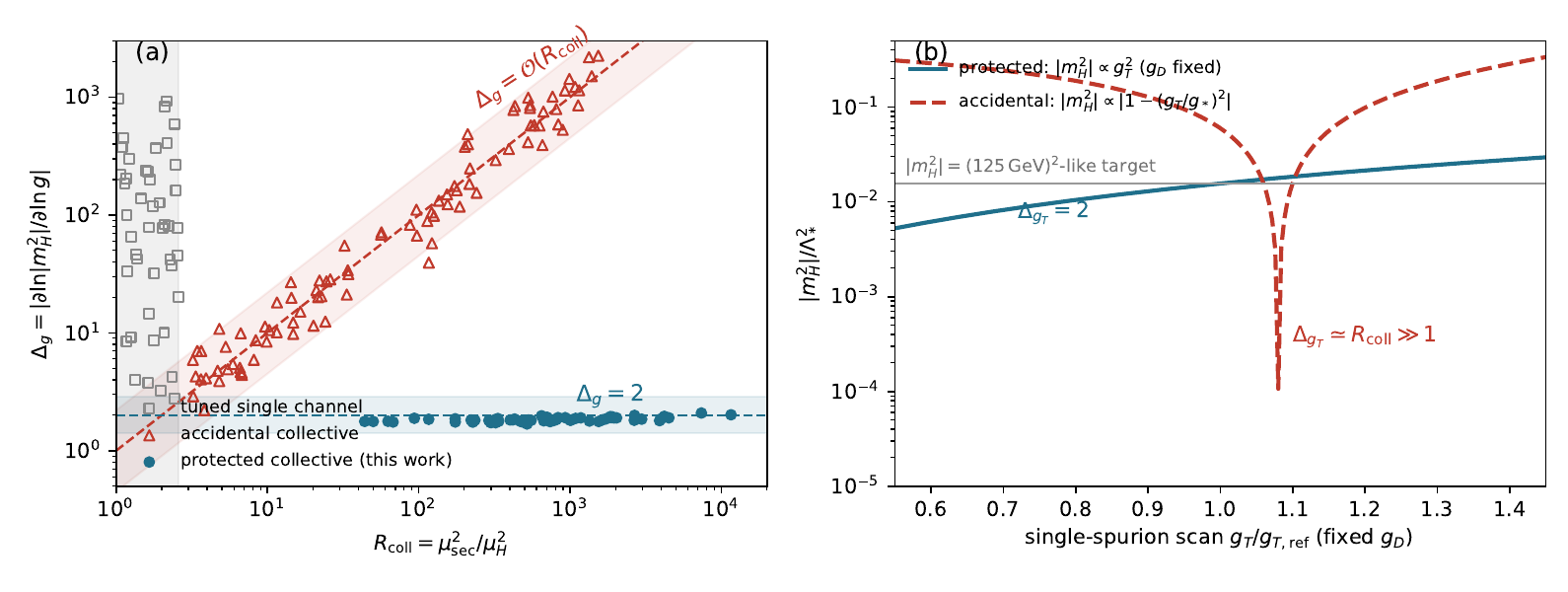}
\caption{Naturalness diagnostics for a light composite scalar. (a)~Tuned
single-channel points have $\Rcoll=\mathcal{O}(1)$; accidental collective
kernels have $\Psec>1$ but $\Delta_g=\mathcal{O}(\Rcoll)$; the protected
collective eigenmode has $\Psec>1$, arbitrarily large $\Rcoll$, and
$\Delta_g=\mathcal{O}(1)$. Blue points: Eqs.~(\ref{eq:C0})--(\ref{eq:mbr})
with $d_X=3$, $f_B=f_H=1$~TeV, $\Lambda_{\rm match}=10$~TeV, $m_X=0.5$~TeV,
$g_T,g_D\in[0.15,0.55]$, $M_\perp\in[0.8,4]$~TeV, with $\Delta_g$ computed
numerically from the full one-loop potential including the threshold
logarithm; the protected points span $\Delta_g\in[1.7,2.1]$. (b)~Response to a
single-spurion scan, $g_T$ varied at fixed $g_D$, with
$\Lambda_*=f_B=1$~TeV. The protected branch follows $|m_H^2|\propto g_T^2$,
whereas an accidental cancellation gives a near-critical dip with local slope
$\mathcal{O}(\Rcoll)$. Rescaling both spurions gives $\Delta_\lambda=4$, still
$\mathcal{O}(1)$.}
\label{fig:trichotomy}
\end{figure*}

Decompose the operator space into sectors $V=\oplus_A V_A$ with projectors
$\Pi_A$ obeying $\sum_A\Pi_A=\mathds{1}$ and $\Pi_A^\dagger Z=Z\Pi_A$. The
canonical projectors $P_A=Z^{1/2}\Pi_A Z^{-1/2}$ are Hermitian and orthogonal.
Define the block-restricted kernel $C_{\rm diag}=\sum_A P_A C P_A$ and
\begin{equation}
\mu_H^2=\lambda_{\min}(C),\qquad
\mu_{\rm sec}^2=\lambda_{\min}(C_{\rm diag}).
\label{eq:muH}
\end{equation}
Since $C_{\rm diag}$ is block diagonal, its minimal eigenvector $w$ lies in a
single block, where $w^\dagger C_{\rm diag}w=w^\dagger Cw\ge\mu_H^2$; hence
$\mu_H^2\le\mu_{\rm sec}^2$ always. For the normalized light eigenvector
$u_H$,
\begin{equation}
\begin{aligned}
\Rcoll&=\frac{\mu_{\rm sec}^2}{\mu_H^2},\qquad
\Psec=\Big(\sum_A p_A^2\Big)^{-1},\\
p_A&=u_H^\dagger P_A u_H.
\end{aligned}
\label{eq:diagnostics}
\end{equation}
The collective spectral regime is
\begin{equation}
\mu_H^2\ll\Lambda_*^2,\quad
\mu_{\rm sec}^2=\mathcal{O}(\Lambda_*^2),\quad
\Rcoll\gg1,\quad \Psec>1.
\label{eq:regime}
\end{equation}
Equation~(\ref{eq:regime}) is basis invariant, but it is only a diagnostic: a
small eigenvalue of a coupled kernel can be produced by an accidental
determinant just as a small single-channel mass can be produced by tuning to
criticality. Naturalness is probed by the response to microscopic bridge
couplings $g_i$~\cite{tHooft:1979rat},
\begin{equation}
\Delta_{g_i}=\left|\frac{\partial\ln|m_H^2|}{\partial\ln g_i}\right|.
\label{eq:Delta}
\end{equation}
When $m_H^2$ arises as an incomplete cancellation between
$\mathcal{O}(\Lambda_*^2)$ entries, individual variations are amplified by the
cancellation ratio, $\Delta_g=\mathcal{O}(\Rcoll)$; a protected collective
eigenmode instead has $\Delta_g=\mathcal{O}(1)$. The resulting trichotomy is
displayed in Fig.~\ref{fig:trichotomy}(a). The remainder of this Letter shows that the protected case is realized
dynamically.

\textit{Rank-one locked zero mode.}---Consider two electroweak-active
composite directions $H_T\sim\delta(\bar Q_TQ_T)$ and
$H_D\sim\delta(\bar\Psi_D\Psi_D)$ with decay constants $F_T$, $F_D$, and
$v^2=F_T^2+F_D^2$. The electroweak-aligned and orthogonal unit vectors are
\begin{equation}
q=\frac{1}{v}\begin{pmatrix}F_T\\F_D\end{pmatrix},\qquad
q_\perp=\frac{1}{v}\begin{pmatrix}-F_D\\F_T\end{pmatrix}.
\label{eq:qvec}
\end{equation}
The protected limit is defined by an enhanced TC--DTC locking symmetry: the
aligned quadratic invariant $|F_TH_T+F_DH_D|^2$ is symmetry-forbidden, while
the orthogonal invariant is allowed,
\begin{equation}
V_{\rm lock}=\frac{M_\perp^2}{2v^2}\,|F_TH_D-F_DH_T|^2,
\label{eq:Vlock}
\end{equation}
equivalently
\begin{equation}
C_0=M_\perp^2\,q_\perp q_\perp^\dagger,\qquad
C_0q=0,\qquad C_0q_\perp=M_\perp^2 q_\perp.
\label{eq:C0}
\end{equation}
At the level of the quadratic kernel this is the statement that $C_0$ is
positive semidefinite with a rank deficiency exactly along $q$---an invariance
of $V_{\rm lock}$ under the constrained shift $(H_T,H_D)\to(H_T,H_D)+\epsilon\,
q$. The distinction from a pNGB coset is structural: no $G/H$ vacuum manifold
is invoked, the light state is a fluctuation of condensates rather than a
coset coordinate, and, as shown below, the loop-level stability of the rank
condition is enforced not by a shift symmetry of the ultraviolet theory but by
a two-spurion collective structure. The aligned scalar
$H=q^\dagger(H_T,H_D)^T$ is an exact zero mode in the protected limit, while
the orthogonal scalar is heavy. At $F_T=F_D$,
$C_0=M_\perp^2\big(\begin{smallmatrix}1&-1\\-1&1\end{smallmatrix}\big)/2$,
giving eigenvalues $0$ and $M_\perp^2$ and $\Psec=2$: the small eigenvalue is
a consequence of a positive-semidefinite invariant, not an entry-by-entry
cancellation.

\textit{DQCD bridge and absence of hard curvature.}---A concrete embedding
uses the dark-technicolor product structure~\cite{Abbas:2020frs,Abbas:2026epjc}
\begin{equation}
G=G_{\rm SM}\times SU(N_{\rm TC})\times SU(N_{\rm DTC})\times SU(N_D),
\label{eq:G}
\end{equation}
where $SU(N_D)$ is the DQCD bridge. The bridge fermions $X_{A,L,R}$, $A=T,D$,
are vectorlike under the gauge symmetries but massless microscopically; DQCD
confinement generates a universal constituent mass $m_X\sim\kappa_X\Lambda_D$,
equivalently a mass function $\Sigma_X(p^2)$ with $\Sigma_X(0)=m_X$. With
bridge projectors $\Pi_T+\Pi_D=\mathds{1}$ and nonlinear link
\begin{equation}
\Sigma(h)=\exp \Big[i\,\frac{h}{f_H}\,\tau_1\Big],
\label{eq:link}
\end{equation}
the matched low-energy mass matrices are
\begin{equation}
\mathcal{M}_A(h)=m_X\mathds{1}
+f_B\big[g_T\Pi_T+g_D\,\Sigma^\dagger(h)\Pi_A\Sigma(h)\big].
\label{eq:MA}
\end{equation}
Here $m_X$ is not a bare vectorlike mass; it is the DQCD-generated constituent
mass. Defining $\mathcal{M}_A^2=\mathcal{M}_A^\dagger\mathcal{M}_A$ and
$P_A(h)=\Sigma^\dagger\Pi_A\Sigma$, projector completeness gives
\begin{equation}
\begin{aligned}
\sum_A\Tr\mathcal{M}_A^2(h)=2\Big[&\,2m_X^2+2m_Xf_B(g_T+g_D)\\
&+f_B^2\big(g_T^2+g_Tg_D+g_D^2\big)\Big],
\end{aligned}
\label{eq:TrM2}
\end{equation}
which is independent of $h$. Therefore
\begin{equation}
\partial_h^2\sum_A\Tr\mathcal{M}_A^2(h)\Big|_0=0.
\label{eq:nohard}
\end{equation}
The hard matched-scale curvature vanishes for arbitrary values of the two
universal locked spurions. If $g_D$ is copy dependent,
\begin{equation}
\partial_h^2\sum_A\Tr\mathcal{M}_A^2\Big|_0
=\frac{4f_B^2g_T}{f_H^2}\Big(g_D^{(D)}-g_D^{(T)}\Big),
\label{eq:nonuniv}
\end{equation}
so non-universality immediately regenerates hard sensitivity. The structure is
rigid in a second way: the $g_T$ spurion is common to both copies and
undressed by $\Sigma(h)$; this relative dressing is what makes $h$ physical
(dressing both spurions renders $h$ a global rotation removable by a field
redefinition), while completeness $\sum_A\Pi_A=\mathds{1}$ of the dressed
projectors removes the hard curvature. With universality, the first
$h$-dependent threshold term is
\begin{equation}
\sum_A\Tr\mathcal{M}_A^4(h)\Big|_{h\text{-dep}}
=-f_B^4g_T^2g_D^2\sin^2 \frac{2h}{f_H},
\label{eq:TrM4}
\end{equation}
\begin{equation}
\partial_h^2\sum_A\Tr\mathcal{M}_A^4\Big|_0
=-\frac{8f_B^4g_T^2g_D^2}{f_H^2}.
\label{eq:TrM4curv}
\end{equation}
The logarithm is the DQCD-matched threshold logarithm, not an arbitrary
elementary cutoff. For $d_X$ Dirac bridge multiplicities, the one-loop
potential~\cite{Coleman:1973jx} gives
\begin{equation}
\begin{aligned}
m^2_{\rm br}&=-\frac{d_X}{2\pi^2}\,g_T^2g_D^2\,\frac{f_B^4}{f_H^2}\,L_X
+\text{finite},\\
L_X&\equiv\ln\frac{\Lambda^2_{\rm match}}{m_X^2}=\mathcal{O}(1\text{--}5).
\end{aligned}
\label{eq:mbr}
\end{equation}
A nonlocal treatment replaces $m_X$ by $\Sigma_X(p^2)$, but
Eq.~(\ref{eq:nohard}) holds pointwise in momentum whenever this mass function
is universal in bridge-copy space. The sensitivity is fixed,
\begin{equation}
\Delta_{g_T}=\Delta_{g_D}=2,
\label{eq:Delta2}
\end{equation}
up to threshold logarithms: the aligned scalar is lifted only at joint spurion
order.For \(m_{\rm br}^2<0\), the aligned direction is destabilized at
\(h=0\); with positive higher-order terms the electroweak-breaking
minimum is generated radiatively. The periodic
potential $\propto-\sin^2(2h/f_H)$ has curvature of equal magnitude at its
minimum, so the protected sensitivity~(\ref{eq:Delta2}) applies to the
physical mass as well.

\textit{Electroweak, top, and oblique viability.}---Because
$m^2_{W,Z}\propto(F_T+H_T)^2+(F_D+H_D)^2$, the electroweak current direction
is $q$. In the protected limit $u_H=q$, so $\kappa_V=1$. An
alignment-breaking perturbation
$\delta C=\epsilon_{\rm al}M_\perp^2(\alpha\,qq_\perp^\dagger+{\rm h.c.})$
gives
\begin{equation}
\kappa_V=1-\tfrac{1}{2}\epsilon_{\rm al}^2|\alpha|^2
+\mathcal{O}(\epsilon_{\rm al}^4),
\label{eq:kappaV}
\end{equation}
so a maximally collective Higgs can remain SM-like in $HVV$ at first order.

The dominant remaining sensitivity is the top sector. The DTC topology
supplies a collective completion: $q_{L3}$ couples through an ETC spurion
$\lambda_L$ to the TC/DQCD side, while $t_R$ couples through an EDTC spurion
$\lambda_R$ to the DTC/DQCD side; DQCD closes the
chain~\cite{Abbas:2020frs,Abbas:2026epjc}. The protected top completion obeys
\begin{equation}
V_{\rm top}(h)=0 \ \ \text{if any of}  \ \
\lambda_L,\lambda_R,g_T,g_D \ \ \text{is switched off}.
\label{eq:Vtop}
\end{equation}
Thus
\begin{equation}
y_t\sim c_t\,\lambda_L\lambda_R\,g_Tg_D,
\label{eq:yt}
\end{equation}
\begin{equation}
\delta m^2_{H,{\rm top}}\sim
-\frac{N_cd_t}{16\pi^2}\,\lambda_L^2\lambda_R^2g_T^2g_D^2f_t^2L_t.
\label{eq:mtop}
\end{equation}
This estimate is parametric, up to threshold factors. A direct single-spurion
Yukawa giving $-N_cy_t^2\Lambda_*^2/(8\pi^2)$ is excluded from the protected
limit.

The scalar protection does not by itself solve the oblique $S$
problem~\cite{Peskin:1990zt}, but the same electroweak-neutral bridge gives a
decoupling route. A hidden-local link~\cite{Bando:1984ej,Bando:1987br}
$\Omega_B$ with
\begin{equation}
\mathcal{L}_B=\frac{f_B^2}{4}\Tr[(D_\mu\Omega_B)^\dagger D^\mu\Omega_B]
\supset\frac{f_B^2}{4}\Tr(g_{\rho T}\rho_{T\mu}-g_B\rho_{B\mu})^2
\label{eq:hidden}
\end{equation}
raises the mostly-TC vector mass without contributing to $v$,
\begin{equation}
M^2_{\rho_T,{\rm eff}}\simeq\frac{g_{\rho T}^2}{4}(F_T^2+f_B^2),\qquad
S_{\rho_T}\simeq\frac{16\pi c_\rho}{g_{\rho T}^2}\,
\frac{F_T^2}{F_T^2+f_B^2}.
\label{eq:Srho}
\end{equation}
For $F_T=v/\sqrt2$, $f_B=f_H=1$~TeV, and $g_{\rho T}=5$, this gives
$M_{\rho T}\simeq2.5$~TeV and $S_{\rho_T}\simeq0.06$ for $c_\rho=1$.

\textit{Lattice falsification.}---The construction gives a direct lattice
target. In a basis $\{\mathcal{O}_T,\mathcal{O}_D,\mathcal{O}_{TD},\dots\}$
solve the generalized eigenvalue problem~\cite{Michael:1985ne,Luscher:1990ck}
\begin{equation}
\mathcal{C}(t)v_n=\lambda_n(t,t_0)\,\mathcal{C}(t_0)v_n.
\label{eq:gevp_lat}
\end{equation}
With $G_0=\mathcal{C}(t_0)$, sector projectors obey
$\Pi_A^\dagger G_0=G_0\Pi_A$, and
\begin{equation}
\begin{aligned}
p_A^{\rm lat}&=v_H^\dagger G_0\Pi_Av_H,\\
P_{\rm sec}^{\rm lat}=\Big[\sum_A(p_A^{\rm lat})^2\Big]^{-1}&,\qquad
R_{\rm lat}=\frac{\min_A(m^{\rm iso}_{0,A})^2}{m_H^2}.
\end{aligned}
\label{eq:lat_diag}
\end{equation}
Protection is tested by a coupling scan,
\begin{equation}
\Delta^{\rm lat}_{g_i}
=\left|\frac{\partial\ln m_H^2}{\partial\ln g_i}\right|,\qquad
\frac{\Delta^{\rm lat}_{g_i}}{R_{\rm lat}}\ll1.
\label{eq:lattice_response}
\end{equation}
Feynman--Hellmann responses~\cite{Bouchard:2016heu} remain
$\mathcal{O}(m_H^2/g_i)$ for the protected branch but are amplified as
$\mathcal{O}(R_{\rm lat})$ for an accidental cancellation. Thus the tuned,
accidental, and protected cases are separable with the same flavor-singlet
correlator and variational machinery already used in BSM lattice
spectroscopy~\cite{Cacciapaglia:2020kgq,Aoki:2014oha,Appelquist:2016viq}.

\textit{Relation to prior work.}---Little-Higgs collective
breaking~\cite{ArkaniHamed:2001nc,ArkaniHamed:2002qy} also requires several
spurions before a scalar mass appears, but the protected object there is a
pNGB shift symmetry of a specified coset; here it is the smallest eigenvalue
of a strongly coupled scalar kernel. Twin Higgs theories~\cite{Chacko:2005pe}
use a partner symmetry to cancel visible quadratic sensitivity, whereas here
the Higgs is a delocalized TC--DTC eigenmode and Eq.~(\ref{eq:nohard}) holds
for arbitrary locked-spurion values. Scalar-seesaw, multisector mixing, and
multi-phase criticality~\cite{Doff:2016jzk,Kannike:2021zzn} can yield
$\Rcoll\gg1$ and $\Psec>1$, but without the response criterion they lie in the
accidental row. Top-sector alternatives without colored
partners~\cite{Bally:2022naz} are complementary to the collective DTC
completion above.

\textit{Conclusion.}---Spectral collectivity and naturalness are logically
independent; the response exponent $\Delta_g$ completes $(\Rcoll,\Psec)$ into
a basis-invariant tuned/accidental/protected classification of light composite
scalars. The protected class is nonempty: a rank-one TC--DTC locking invariant
removes tree-level aligned curvature, a universal DQCD bridge removes hard
one-loop curvature, and any bridge non-universality regenerates it. The same
DTC topology supplies a collective top completion and a vector-decoupling
route for $S$, while the trichotomy is testable by sector-restricted GEVP
spectroscopy and coupling-response scans. The protected object is not a
coordinate of a coset; it is an eigenvalue of a strongly coupled scalar
kernel, and it is measurably so.

\section*{Acknowledgment}
 
%\begin{appendix}
%
%\end{appendix}

\end{document}